\documentclass [a4paper,10pt]{article}
\usepackage{amsfonts,amsmath,amssymb}
\begin{document}

\title{ On two-point correlation functions in AdS/QCD }
\author{ A. Krikun \thanks{e-mail: krikun.a@gmail.com}\\ \textit{MIPT and ITEP}, \textit{Russia, Moscow}}
\date{}

\maketitle

\begin{abstract}
In this paper we study the chiral symmetry breaking in the hard wall AdS/QCD model. We solve the equations of motion up to the second order at large momentum and obtain the first few terms in the expansion of the "left-right" correlator, which is the measure of symmetry breaking. We study the dependence on AdS radius to get the result as the series in t'Hooft constant.
\end{abstract}

\section {Introduction}

\indent
In the last few years a great attention was paid to the so-called phenomenological AdS/QCD theories. The essence of these models is to use the AdS/CFT correspondence \cite{Maldacena} to describe QCD in large $N_c$ limit via its 5-dimensional dual theory.
The exact structure of this 5D theory, describing all specific features of QCD is not clear, but some simple models have been proposed \cite{Erlich} \cite{Erlich+} \cite{Pomarol1} \cite{Pomarol2} \cite{Evans1} \cite{Resh}, which already give promising results.

In this paper we study the simplest of these settings, the so-called hard wall AdS/QCD model (see for example \cite{Erlich} \cite{Pomarol1}, first proposed in \cite{HW}) . Our goal is to find solutions to the equations of motion in 5D theory perturbatively in inverse powers of momenta and analyze two-point correlators at large $Q^{2}$ and their dependence on AdS curvature radius. Classical solutions play an important role in AdS/QCD, because, according to AdS/CFT correspondence, correlation functions in QCD can be evaluated as variations of 5D effective action on the classical trajectories with respect to the boundary values of the corresponding fields.
$$
Z^{4D}[J_{1}(x),J_{2}(x),...]=exp\left(S^{5D}_{eff}[V_{1}(z,x)_{classic},V_{2}(z,x)_{classic},...]\right)|_{V_{i}(0,x)=J_{i}(x)}
$$

This rule allows us to compute 2-point functions for vector, axial-vector and pseudoscalar  currents in QCD, fix the free parameters of the model and also study the so-called "left-right" correlator ($\langle LR \rangle=\langle VV \rangle- \langle AA \rangle $) up to the term, quadratic in condensate.
The dependence on QCD coupling constant is reconstructed in the result, via the ADS/CFT correspondence of Yang-Mills coupling constant and AdS curvature radius (see for review \cite{AdS/CFT}).
$$
\frac{R^{4}}{4 \pi \alpha'^{2}}= \alpha_{s} N_{c}=\lambda'=N_c g_{ym}^2
$$

We obtain the expression for the "left-right" correlator at the strong coupling regime. The first term, contrary to the sum rules result \cite{sumrule}, which is linear in $\lambda'$, is proportional to $\lambda'^{0}$. This is not strange, because calculations via AdS/QCD imply the large coupling constant, while sum rules work well at weak coupling.

This paper is organized as follows. In Section 2 we give a review of AdS/QCD hard wall model and derive action for axial, vector and pseudoscalar fields and corresponding equations of motion. In Section 3 we solve these equations in the limit of large momenta by means of the Green function, derived in Appendix. Section 4 is dedicated to the evaluation of the correlation functions via the AdS/CFT recipe and matching of parameters in result with QCD ones. Conclusion is given in the Section 5.

\section{The description of the model}
\indent
Here we consider the simplest holographic model of low energy QCD, proposed in \cite{Erlich},\cite{Pomarol1},\cite{Evans1}(see also \cite{QCD1,QCD2,QCD3}), the \hbox{so-called} "Hard wall AdS/QCD model". In what follows we will work with conventions and notations, used in \cite{Erlich}.

In the AdS/CFT prescription, the fields in 5-dimensional space are dual to operators in 4D, and the global flavor symmetry of the 4D field theory corresponds to the gauge symmetry in its 5D dual. So we will study 4D QCD with $SU(2)_L \times SU(2)_R$ global symmetry via the gauge theory in AdS with $SU(2)_L \times SU(2)_R$ gauge group. In this model only the fields, dual to QCD operators with the lowest dimensions, are considered.

\subsection {The metric and the fields}

 We have the $SU(2)_L \times SU(2)_R$ gauge field theory in $AdS_5$ space with the metric:
\begin{equation}
\label{1}
ds^{2}=\frac{R^2}{z^{2}}(-dz^{2} + dx^\mu dx_\mu)
\end{equation}
where $R$ is the AdS curvature radius, cut at $z$ coordinate: $0<z\leq z_m$

Later, we will denote 5-dimensional indices with capital Latin letters {$(A,B..\in\{0,1,2,3,z\})$}, and 4D indices with Greek letters $(\mu,\nu,..\in\{0,1,2,3\})$.
We will use the metric tensors, defined as follows:
\begin{gather*}
g_{AB} = diag \left (-\frac{R^{2}}{z^{2}},\frac{R^{2}}{z^{2}},-\frac{R^{2}}{z^{2}},-\frac{R^{2}}{z^{2}},-\frac{R^{2}}{z^{2}} \right)  \qquad  \\
\eta_{AB} = diag(-1,1,-1,-1,-1) \\
g=det(g_{AB}) = \frac{R^{10}}{z^{10}}\\
\end{gather*}

The theory includes left- and right-handed gauge vector fields of $SU_L(2)~ \times ~SU_R (2)$ ($A_L$ and $A_R$, respectively) and bifundamental scalar $X_{\alpha \beta}$. According to AdS/CFT 5D fields correspond to operators in~QCD:
\begin{align*}
A_{L \mu}^{a} &\leftrightarrow \bar{q}_L \gamma^{\mu} t^{a} q_L \\
A_{R \mu}^{a} &\leftrightarrow \bar{q}_R \gamma^{\mu} t^{a} q_R \\
\left ( \frac{2}{z} \right ) X^{\alpha \beta} &\leftrightarrow \bar{q}_R^{\alpha} q_L^{\beta}
\end{align*}
with the boundary conditions imposed at $z=z_m$:
$$
\partial_z V(z_m) =0 \ \ ; \ \ \partial_z A(z_m) =0.
$$

\noindent The action is:
\begin{equation}
\label{2}
S=\int d^{5}x \sqrt{g} Tr \left \{ \Lambda^2(|DX|^{2} + \frac{3}{R^{2}}|X|^{2}) - \frac{1}{4 g_5^{2}} (F_L^{2} + F_R^{2})\right \}
\end{equation}
where
$$
D_B X= \partial_B X - \imath A_{L B}X + \imath X A_{R B}
$$
$$
A_{L(R)} = A_{L(R)}^{a} t^{a}
$$
$$
F_{B D}=\partial_B A_D - \partial_D A_B - \imath[A_B , A_D],
$$
and we introduce the normalization constant $\Lambda$ of field $X$.

\noindent From lagrangian for X in absence of $A$ and $V$ (\ref{2}), we get the equation of motion:
\begin{gather*}
\frac{1}{z^{5}} 3 X = \frac{1}{z^{3}} \partial_{\mu} \partial^{\mu} X - \partial_z \frac{1}{z^{3}} \partial_z X
\end{gather*}
which has the solution:
$$
X_0 (z) = \frac{1}{2} M z + \frac{1}{2} \Sigma z^{3}.
$$
According to AdS/CFT \cite{AdS/CFT}, we argue, that $M$ corresponds to quark mass matrix, i.e. the source of operator $\bar{q}_R^{\alpha} q_L^{\beta}$ and $\Sigma$ to condensates, i.e. vacuum expectation value of $\bar{q}_R^{\alpha} q_L^{\beta}$. We can make $M$ to be quark masses exactly by appropriate definition of normalization $\Lambda$, but the relation between $\Sigma$ and condensates is to be ascertained. In further discussion we set $\Sigma=\sigma \textbf{1}$, $M = m \textbf{1}$, assuming the equality of quark masses.
\begin{equation}
\label{3}
X_0 (z) = \frac{1}{2} v(z) \textbf{1}, \qquad v(z) = mz+\sigma z^{3}
\end{equation}

\subsection{The action for vector and axial vector fields}

It is convenient to rewrite the action in terms of vector and axial vector fields:
\begin{align*}
V &=(A_L+A_R)/2 \\
A &=(A_L-A_R)/2.
\end{align*}

To get the effective action for small fluctuations of fields, we will derive it up to quadratic order.
$$
F_L^2=F_{L,AB}F_L^{AB}=F_{L,AB}F_{L,CD} \bigl(\frac{z^{2}}{R^2} \eta^{AC}\bigr)\bigl(\frac{z^{2}}{R^2}\eta^{BD}\bigr)
$$
\begin{align*}
F_{L,AB}F_{L,CD}&= (\partial_A V_B + \partial_A A_B - \partial_B V_A - \partial_B A_A )\cdot(\partial_C V_D + \partial_C A_D - \partial_D V_C - \partial_D A_C )=\\
&= F_{V,AB}F_{V,CD} + F_{A,AB}F_{A,CD} +F_{V,AB}F_{A,CD} + F_{A,AB}F_{V,CD}\\
 F_{R,AB}F_{R,CD}&= F_{V,AB}F_{V,CD} + F_{A,AB}F_{A,CD} - F_{V,AB}F_{A,CD} - F_{A,AB}F_{V,CD}
\end{align*}

\noindent As can be seen, the cross-terms cancel, so the gauge part of (\ref{2}) takes form:
\begin{align}
S_{AA,VV} &= \int d^{5} x \frac{R^5}{z^{5}} \left ( - \frac{1}{4 g_5^{2}} \right ) \bigl(\frac{z^{2}}{R^2} \eta^{AC}\bigr)\bigl(\frac{z^{2}}{R^2} \eta^{BD}\bigr) 2( F_{A,AB}^{a} F_{A,CD}^{b} + F_{V,AB}^{a} F_{V,CD}^{b} ) Tr \{ t^{a} t^{b} \}=\notag\\
\label{4}
&=\int d^{5} x \left ( - \frac{R}{4 g_5^{2}} \right ) \frac{1}{z} ( F_{A}^{a} F_{A}^{a} + F_{V}^{a} F_{V}^{a} )
\end{align}
where we introduced the notation: $\eta^{AC} \eta^{BD} F_{A,AB}^{a} F_{A,CD}^{a}=F_{A}^{a} F_{A}^{a}$.

Note, that in $F_{A}^{2}$ and $F_{V}^{2}$ we have the same terms with odd number of $A$, arising with opposite signs. So they will cancel each other and there will be no terms in result, containing odd number of \hbox{axial fields $A$} \hbox{(for example $AVV$)}.\\

Now we consider the part of the action (\ref{2}), describing the interaction of $X$ with the gauge fields and decompose $X$ as follows:
$$
X^{\alpha\beta} = X_{0}^{\alpha\gamma} exp ( \imath 2 \pi^{a} (t^{a})^{\gamma\beta} ) , \qquad X_{0}^{\alpha\gamma} = \delta^{\alpha\gamma}\cdot \frac{1}{2}(mz+\sigma z^3) =\delta^{\alpha\gamma}\cdot \frac{1}{2} v(z).
$$
Expanding $X$ to the first order in $\pi=\pi^{a}t^{a}$: $X=X_0(1+2\imath \pi)$ we get
\begin{align*}
D_A X &= X_{0} ( 2\imath \partial_A \pi - \imath L_A + 2L_A \pi + \imath R_A - 2\pi R_A )\\
D_A X D_B X &= v(z)^{2} (\partial \pi - A)_A (\partial \pi - A)_B
\end{align*}
therefore
\begin{align}
S_{int} &= \int d^{5}x \sqrt{g} Tr \left \{\Lambda^2 |DX|^{2} \right \} \\
 &= \int d^{5}x \frac{R^5}{z^{5}} ( \frac{z^{2}}{R^2} \eta^{AB} ) \Lambda^2 v(z)^{2} (\partial \pi^{a} - A^{a})_A (\partial \pi^{b} - A^{b})_B Tr\{ t^{a} t^{b} \}:=\notag\\
\label{5}
&:=\int d^{5}x \frac{R^{3} \Lambda^2 v(z)^{2}}{2z^{3}} (\partial \pi - A)^{2}.
\end{align}

\subsection{Equations of motion}
\indent
We have obtained the action in terms of $V,A$ and $\pi$. Now, let us derive the equations of motion, imposing the gauge \cite{Erlich}:
\begin{equation}
\label{5.5}
A_z = V_z =0
\end{equation}
$$
\partial_{\mu} V_{\mu} = 0
$$
$$
A_{\mu}=A_{\perp \mu} + \partial_{\mu}\phi \ \; \ \
\partial_{\mu} A_{\perp \mu} = 0.
$$
For $V_A$ we get (\ref{4}):
\begin{align*}
- \partial_z \frac{1}{z} F_{zB} + \frac{1}{z} \partial_{\mu} \eta^{\mu \nu} F_{\nu B} &=0 \\
-\partial_z \frac{1}{z} \partial_z V_{\mu} + \frac{1}{z} \triangle V_{\mu} &=0
\end{align*}
and after the Fourier transform \hbox{$V(q,z)=\int d^{4} x \ \ e^{\imath q x} V(x,z)$} it takes form
\begin{equation}
\label{6}
\partial_z \frac{1}{z} \partial_z V_{\mu}^{a} (q,z) + \frac{q^{2}}{z} V_{\mu}^{a}(q,z) =0.
\end{equation}

The action of axial sector in terms of $A_{\perp \mu}$, $A_z$, $\phi$ and $\pi$ is (\ref{4}),(\ref{5}):

\begin{align}
S_{A} =\int d^{5} x \left ( - \frac{R}{4 g_5^{2}} \right ) \frac{1}{z} ( F_{A \perp} F_{A \perp}) + \left (  \frac{R}{2 g_5^{2}} \right ) \frac{1}{z} (\partial_{z}\partial_{\mu}\phi - \partial_{\mu} A_z)^{2} + \notag\\
\label{S_A}
+ \frac{\Lambda^2 R^3 v(z)^{2}}{2  z^3} \left[ -( \partial_{z} \pi - A_z)^{2} + (\partial_{\mu} \pi -\partial_{\mu}\phi )^{2} + (A_{\perp \mu})^2\right]
\end{align}
and gives the following equations of motion (the gauge $A_z=0$ is used):
\begin{align}
\label{7}
\left [ \partial_z \left ( \frac{1}{z} \partial_z A_{\mu}^{a} \right ) + \frac{q^2}{z} A_{\mu}^{a} - \frac{R^{2} g_{5}^{2} \Lambda^2 v^{2}}{z^{3}} A_{\mu}^{a} \right ]_\perp &= 0\qquad \mbox{for $A_{\perp \mu}$,} \\
\label{8}
\partial_z \left ( \frac{1}{z} \partial_z \phi^{a} \right ) + \frac{R^{2} g_{5}^{2} \Lambda^2 v^{2}}{z^{3}} (\pi^{a}- \phi^{a})&= 0 \qquad \mbox{for $\partial_{\mu} \phi$,}\\
\label{9}
-q^2 \partial_z \phi^{a} + \frac{R^{2} g_{5}^{2} \Lambda^2 v^{2}}{z^{2}} \partial_z \pi^{a} &= 0 \qquad \mbox{for $A_z$,}\\
\label{9.5}
\partial_{z} \frac{v^2}{z^3} \partial_{z} \pi + q^{2} \frac{v^2}{z^3} (\pi - \phi) &= 0 \qquad \mbox{for $\pi$.}
\end{align}
One can see, that (\ref{8}) follows from (\ref{9}) and (\ref{9.5}), so we can use the last two to find solutions for $\phi$ and $\pi$.

\section {Solution to the equations of motion}

According to AdS/CFT correspondence, the generating functional of correlation functions in 4D conformal theory equals to the effective action of its dual 5D theory computed on the classical trajectories \cite{Maldacena}\cite{AdS/CFT}. Correlation functions can be obtained by the variation of the 5D action with respect to the boundary values of fields. For example, vector current correlator is:
$$
\langle J_V (q_{1}) J_V (q_{2}) \rangle = \frac {\delta}{\delta V_0(q_{1})}\frac {\delta}{\delta V_0(q_{2})} S(V_0)|_{V_0=0}.
$$
It is convenient to introduce functions $v(q,z)$,$a(q,z)$ and $\phi(q,z)$, such that:
\begin{align}
\label{10}
&V_A(q,z)=V_{0A} (q) v(q,z) \ \ ; \ \ v(q,z)|_{z=\epsilon} = 1 \notag \\
&A_A(q,z)=A_{0A} (q) a(q,z) \ \ ; \ \ a(q,z)|_{z=\epsilon} = 1 \notag \\
&\phi(q,z)|_{z=\epsilon} = \phi_{0} (q)
\end{align}
and use the euclidian momentum $Q^2=-q^2$.

\subsection{Solution for $V$}

The equation for the function $v$ (\ref{6})(\ref{10}) is:
$$
\partial_z \frac{1}{z} \partial_z v - \frac{Q^{2}}{z} v =0
$$
which can be reduced to:
$$
u'' +\frac{u'}{z} + u \left ( -Q^{2} - \frac{1}{z^{2}} \right ) =0,\qquad u=\frac{v}{z}.
$$
This is the modified Bessel equation with $\lambda = 1 $. Its solution is:
$$
u(z)=const(A \textit{I}_{1}(Qz) + B \textit{K}_{1}(Qz))
$$
where $\textit{I}_{1}$ and $\textit{K}_{1}$ - are modified Bessel functions of the first and second kind respectively.
Constants in this solution can be fixed, using the boundary conditions on $v(z)$ at $z=\epsilon$ (\ref{10}) and $z=z_m$~(\ref{5.5}),
$$
A=\textit{K}_{0}(Qz_m) \qquad B= -\textit{I}_{0}(qz_m)
$$
$$
const = \frac{Q}{B}
$$
and we obtain the solution for $v$:
\begin{equation}
\label{11}
v(Q,z)=- \frac{1 }{\textit{I}_{0}(Qz_m)} Qz[ \textit{K}_{0}(Qz_m)\textit{I}_{1}(Qz) - \textit{I}_{0}(Qz_m) \textit{K}_{1}(Qz) ] = \frac{1}{B} x ( A \textit{I}_{1}(x) + B \textit{K}_{1}(x) ).
\end{equation}

\subsection{Solution for $A$}

The equation for $A_{\perp}$ (\ref{7}) differs from the equation for $V$ (\ref{6}) by the interaction term. The solution can't be found analytically, but we can study its asymptotic at large euclidian momentum $Qz_m \gg 1$ in the chiral limit $m^2 \rightarrow 0$. We will also keep the $m \sigma$ term, to study the condensate dependence of the solution.

\begin{align*}
\partial_z \left ( \frac{1}{z} \partial_z a \right ) - \frac{Q^2}{z} a &= \frac{R^{2} g_{5}^{2} \Lambda^2 v^{2}}{z^{3}} a  \\
-\frac{1}{z} \partial_z a + \partial_{z}^{2} a - Q^{2} a &= R^{2} g_{5}^{2} \Lambda^2 \sigma^{ 2} z^{4} a + R^{2} g_{5}^{2} \Lambda^2 (2m \sigma) z^{2} a
\end{align*}
Changing the variable $Qz\rightarrow x$, we get at large $Q$ the inhomogeneous Bessel equation ($\lambda = 1$) for function $xa(x)$ (for convenience we denote small parameters $\lambda = \frac{g_{5}^{2} \Lambda^2 \sigma^{ 2} R^2}{Q^{6}}$ and $\mu = \frac{g_{5}^{2} \Lambda^2 (2\sigma m) R^2}{Q^{4}}$ ):
\begin{equation}
\label{11.5}
 -\frac{1}{x} \partial_x a + \partial_{x}^{2} a - a = \lambda x^{4}a + \mu x^{2} a
\end{equation}
It can be solved by means of the Green function, derived in \cite{Resh}, which for the euclidian momentum is presented in Appendix.

We will compute first and second order corrections due to $\lambda$-term, because they don't vanish in the chiral limit and derive $\mu$-term, related to quark mass.

\subsubsection{First order correction}

The first order correction to the solution to the homogeneous equation can be computed, using the Green function (\ref{A1}),  as the integral ($x_0=Q \epsilon$, $x_m=Q z_m$):
$$
a^{(1)} = \int\limits_{x_0}^{x_m} dx' \lambda x'^{4} a^{(0)}(x') \frac{G(x,x')}{x'}
$$
where $a^{(0)}$ - is the solution to the homogeneous equation (\ref{11}), found in the previous Section.

\noindent Inserting the Green function (\ref{A1}), we have:
\begin{align}
a^{(1)}(x)&= \int\limits_{0}^{x_m} dx' \lambda x'^{3} a^{(0)}(x')\cdot G(x,x')\notag\\
\label{12}
&=\frac{x [ A \textit{I}_{1}(x) + B \textit{K}_{1}(x)] }{AD-BC} \int_{0}^{x} dx' \lambda x'^{3} a^{(0)}(x') \cdot x' [ C \textit{I}_{1}(x') + D \textit{K}_{1}(x')] + \notag \\
&+ \frac{x [ C \textit{I}_{1}(x) + D \textit{K}_{1}(x)] }{AD-BC} \int_{x}^{x_m} dx' \lambda x'^{3} a^{(0)}(x')\cdot x' [ A \textit{I}_{1}(x') + B \textit{K}_{1}(x')].
\end{align}
We are mainly interested in the behavior of $a(z)$ at the vicinity of the boundary $z \rightarrow 0$ (consequently $x\rightarrow 0$), hence the major contribution is due to the second integral:
$$
a^{(1)} (x) = \frac{x [ C \textit{I}_{1}(x) + D \textit{K}_{1}(x)] }{AD-BC} \int_{0}^{x_m} dx' \lambda x'^{3} a^{(0)}(x') \cdot B a^{(0)}(x').
$$
In the limit $x_0=q\epsilon \rightarrow 0 $ and $x\rightarrow 0$ we get:
$$
AD-BC = [ A \textit{I}_{1}(x_0)-B \textit{K}_{1}(x_0) ]|_{x_0 \rightarrow 0 } =
A \frac {x_0}{2} - B \frac{1}{x_0} \simeq - \frac{B}{x_0}
$$
$$
x [ C \textit{I}_{1}(x) + D \textit{K}_{1}(x)]|_{x_0 \rightarrow 0} = x \frac{1}{x_0} \textit{I}_{1}(x) |_{x\rightarrow 0} =
 \frac{x^{2}}{2 x_0}
$$
therefore
\begin{align*}
a^{(1)}(x)&= - \frac{x^{2}}{2B}  B \lambda \int\limits_{0}^{x_m} dx' x'^{3} [a^{(0)}(x')]^{2}.
\end{align*}

\noindent In the limit $x_m \rightarrow \infty$,  $a^{(0)}(x)$ (\ref{11}) takes form:
\begin{align}
a^{(0)} (x)
\label{13}
&= -x \sqrt{2 \pi x_m} e^{-x_m}  \left( \sqrt{\frac{\pi}{2x_m}} e^{-x_m} \cdot \textit{I}_{1}(x) - \sqrt{\frac{1}{2 \pi x_m}} e^{x_m} \cdot \textit{K}_{1}(x) \right)=x \textit{K}_{1}(x)
\end{align}
and the integral equals to:
$$
\int\limits_{0}^{\infty} dx' x'^{3} [a^{(0)}(x')]^{2} = \int\limits_{0}^{\infty} dx' x'^{5} [\textit{K}_{1}(x')]^{2}=\frac{8}{5}.
$$

\noindent Consequently we have the first order correction to the solution (near the boundary):
\begin{equation}
\label{14}
a^{(1)}(Q,z) = - \frac{4}{5} (Qz)^{2} \frac{g_{5}^{2} R^2 \Lambda^2 \sigma^{2} }{Q^{6}}.
\end{equation}

\subsubsection{Second order correction}

To compute the second order correction we have to evaluate integral:
$$
a^{(2)} = \int\limits_{0}^{x_m} dx' \lambda x'^{4} a^{(1)} (x') \frac{G(x,x')}{x'}.
$$

\noindent Inserting the Green function (\ref{A1}), we get:
\begin{align*}
a^{(2)}(x)&= \int\limits_{0}^{x_m} dx' \lambda x'^{3} a^{(1)} G(x,x')=\\
&=\frac{x [ A \textit{I}_{1}(x) + B \textit{K}_{1}(x)] }{AD-BC} \int\limits_{0}^{x} dx' \lambda x'^{3} a^{(1)} x' [ C \textit{I}_{1}(x') + D \textit{K}_{1}(x')] +\\
&+ \frac{x [ C \textit{I}_{1}(x) + D \textit{K}_{1}(x)] }{AD-BC} \int\limits_{x}^{x_m} dx' \lambda x'^{3} a^{(1)} x' [ A \textit{I}_{1}(x') + B \textit{K}_{1}(x')].
\end{align*}
Similarly to the first correction calculation, we are mainly interested in the behavior of $a(z)$ near the boundary $z \rightarrow 0$ (and $x\rightarrow 0$), so the major input is due to the second integral.
We see, that to compute this integral we have to know the form of the first correction at any x (not only near the boundary, where we've found it (\ref{14})).Therefore we can try to find the form of the function $a^{(2)} (x)$ (and its order on $Q$ and $z_m$) near the boundary and evaluate the coefficient numerically.

In the limits $x_m\rightarrow \infty$ and $x_0\rightarrow 0$, the coefficients A,B,C,D behave as:
\begin{align*}
A|_{x_m\rightarrow \infty}&=\textit{K}_{0}(x_m)|_{x_m\rightarrow \infty}=\sqrt{\frac{\pi}{2x_m}} e^{-x} \rightarrow 0\\
B|_{x_m\rightarrow \infty}&=- \textit{I}_{0}(x_m)|_{x_m\rightarrow \infty}=\frac{1}{\sqrt{2 \pi x_m}} e^{x_m} \\
C|_{x_0\rightarrow 0}&=\textit{K}_{1}(x_0)|_{x_0\rightarrow 0}=\frac{1}{x_0}\\
D|_{x_0\rightarrow 0}&=-\textit{I}_{1}(x_0)|_{x_0\rightarrow 0}=\frac{x_0}{2} \rightarrow 0
\end{align*}
so $a^{(1)}$ (\ref{12}) takes form (\ref{13}):
\begin{align*}
a^{(1)}(x)&= x  \textit{K}_{1}(x) \int\limits_{0}^{x} dx' \lambda x'^{5}  \textit{K}_{1}(x') \textit{I}_{1}(x') + x   \textit{I}_{1}(x)  \int\limits_{x}^{x_m} dx' \lambda x'^{5} \textit{K}_{1}(x')    \textit{K}_{1}(x')
\end{align*}
and we can rewrite $a^{(2)}(x)$ as:
\begin{equation}
\label{15}
 a^{(2)} (x)= \frac{\alpha x^2}{2} \lambda^{2}
\end{equation}
where $\alpha$ is the numerical coefficient:
\begin{align*}
\alpha = \int\limits_{0}^{x_m} dx' x'^{5} \textit{K}_{1}(x') &\left[\textit{K}_{1}(x') \int\limits_{0}^{x'} dx'' x''^{5}  \textit{K}_{1}(x'') \textit{I}_{1}(x'') + \textit{I}_{1}(x') \int\limits_{x'}^{x_m} dx'' x''^{5} \textit{K}_{1}(x'')    \textit{K}_{1}(x'')   \right] \approx \\ &\approx 124.
\end{align*}

\subsubsection{Quark mass correction}
The $\mu$-term correction can be easily computed similarly the subsection \textbf{3.2.1} via the integral:
$$
a_m^{(1)} = \int\limits_{0}^{x_m} dx' \mu x'^{2} a^{(0)}(x') \frac{G(x,x')}{x'}
$$

\noindent and after inserting the Green function and taking the appropriate limit:
\begin{align*}
a_m^{(1)}(x)&= - \frac{x^{2}}{2B}  B \mu \int\limits_{0}^{\infty} dx' x' [a^{(0)}(x')]^{2}.
\end{align*}

\noindent The integral equals to:
$$
\int\limits_{0}^{\infty} dx' x' [a^{(0)}(x')]^{2} = \int\limits_{0}^{\infty} dx' x'^{3} [\textit{K}_{1}(x')]^{2}=\frac{2}{3}
$$
and we get:
\begin{align}
\label{mass}
a_m^{(1)}(x)&= - x^2 \frac{1}{3} \frac{g_{5}^{2} \Lambda^2 (2\sigma m_q) R^2}{Q^{4}}.
\end{align}

Finally, from (\ref{11},\ref{14},\ref{15},\ref{mass}) we obtain the solution to the equation of motion for $a(z)$ up to the second order in $\lambda$ at the vicinity of the boundary ($z\rightarrow 0$) at large $Q^{2}$:
\begin{align}
a(z)&= \frac{1}{\textit{I}_{0}(Qz_m)} Qz [ \textit{K}_{0}(Qz_m)\textit{I}_{1}(Qz) - \textit{I}_{0}(Qz_m) \textit{K}_{1}(Qz)] - \notag \\
\label{a}
&- \frac{4}{5} z^{2} \frac{g_{5}^{2} \Lambda^2 \sigma^{2} R^2}{Q ^{4}}+ 62 z^{2}\frac{g_{5}^{4} \Lambda^4 \sigma^{4} R^4}{Q ^{10}}- z^2 \frac{1}{3} \frac{g_{5}^{2} \Lambda^2 (2\sigma m) R^2}{Q^{2}}.
\end{align}

\subsection{Solution for $\phi$ and $\pi$}
Now we will compute to the leading order solutions for the pseudoscalar fields in the theory. These are solutions to the equations (\ref{9}),(\ref{9.5}) with fixed boundary value of $\phi$ at $z=\epsilon$ (\ref{10}). Differentiating (\ref{9.5}) and substituting $\partial_z \phi$ from (\ref{9}) we get the equation:
$$
\partial_{z}^{2} \frac{v^2}{z^3} \partial_{z} \pi - \left(\partial_{z} \frac{v^2}{z^3}\right) \frac{z^3}{v^2} \partial_{z}\frac{v^2}{z^3} \partial_{z} \pi - Q^2 \frac{v^2}{z^3} \partial_{z} \pi - \frac{g_{5}^{2} R^2 \Lambda^2 v^4}{z^5} \partial_{z} \pi = 0.
$$
We need to solve it near the boundary, so we can substitute for $v(z)$ its asymptotic $v(z)=mz|_{z\rightarrow 0}$ (\ref{3}) and after change $x=Qz$ it takes form:
$$
\partial_{x}^{2} \frac{1}{x} \partial_{x} \pi + \frac{1}{x} \partial_{x} \frac{1}{x} \partial_{x} \pi - \frac{1}{x} \partial_{x} \pi - \frac{g_{5}^{2} R^2 \Lambda^2 m^2}{Q^2} \frac{1}{x} \partial_{x} \pi = 0.
$$
In the large $Q^2$ limit we can neglect the last term and obtain the modified Bessel equation with $\lambda=0$ for the function $\frac{1}{x} \partial_x \pi$. Hence the solution for $\pi(z)$ is:
$$
\pi(z) = A' Q z I_{1}(Qz) + B' Q z K_{1} (Qz).
$$
Then, using (\ref{9}) we immediately obtain the solution for $\phi$:
$$
\phi(z) = - \frac{g_{5}^{2} R^2 \Lambda^2 m^2}{Q^2} Qz [A' I_{1}(Qz) + B' K_{1} (Qz)].
$$
The boundary condition on $\phi$ at $z=\epsilon$ (\ref{10}) fixes the constant $B'$:
$$
B'=-\frac{Q^2}{g_{5}^{2} R^2 \Lambda^2 m^2} \phi_{0} (q)
$$
and we finally get solutions at $z\rightarrow 0$:
\begin{align}
\label{phi_pi}
\phi(z)|_{z=\epsilon}=\phi_{0} (q) \qquad \left. \frac{\partial_{z} \phi(z)}{z} \right|_{z=\epsilon}&= -\phi_{0} (q) \frac{Q^2}{2} ln(Q^2 \epsilon^2) \notag\\
\pi(z)|_{z=\epsilon}= -\phi_{0} (q) \frac{Q^2}{g_{5}^{2} R^2 \Lambda^2 m^2} \qquad \left. \frac{\partial_{z} \pi(z)}{z} \right|_{z=\epsilon}&= \phi_{0} (q) \frac{Q^2}{g_{5}^{2} R^2 \Lambda^2 m^2} \frac{Q^2}{2} ln(Q^2 \epsilon^2).
\end{align}

\section{Computation of correlators}

It was already mentioned in the previous Section, that two-point correlation function can be obtained by the variation of the effective action with respect to the boundary values of fields.

\subsection {$\langle J_V,J_V \rangle$ correlator}

For the beginning let us evaluate the variation of action on the classical solution:
\begin{align*}
\delta S = \int d^{5} x \frac{\delta L}{\delta V} \delta V + \frac{\delta L}{\delta \partial_M V} \delta \partial_M V = \int d^{4} x dz \left[ \frac{\delta L}{\delta V} - \partial_M \frac{\delta L}{\delta \partial_M V}  \right]\delta  V -  \left. \int d^{4} x \frac{\delta L}{\delta \partial_z V} \delta  V \right|_{\epsilon}^{z_m}.
\end{align*}
Hence on the classical solution the variation of action reduces to the boundary term and we get for $V_{\mu}$:
\begin{align*}
\delta S_{V} =   - \int d^{4} x \frac{R}{g_{5}^{2}} \left[ \delta V_{\mu} \frac{\partial_z  V_{\mu}}{z} \right]_{z=\epsilon} = \left ( - \frac{R}{g_5^{2}} \right ) \int d^{4} x d^{4} q_{1} d^{4} q_{2} e^{ \imath q_{1} x} e^{ \imath q_{2} x} \left [ \delta V_{\mu}(q_{1},z) \frac{ \partial_z V_{\mu}(q_{2},z)}{z} \right ]_{z=\epsilon}.
\end{align*}
Taking twice the variation with respect to the boundary value $V^{a}_{0 \mu}(q)$ (\ref{10}) we obtain the current correlator:
\begin{align}
 \langle J_{V \mu}^{a} (q) J_{V \nu}^{b} (q) \rangle = \delta^{ab} (q_{\mu}q_{\nu} - q^{2} g_{\mu \nu}) \frac{R}{g_{5}^{2} q^{2}}
\label{17}
\left [ v(q,z)\frac{\partial_z v(q,z)}{z} \right ]_{z=\epsilon} =\delta^{ab} (q_{\mu}q_{\nu} - q^{2} g_{\mu \nu}) \Pi_V (q^{2})
\end{align}
and for $\Pi_{V}(q^{2})$ we have, using the solution for $v$ (\ref{11}):
\begin{align*}
\left [ \frac{\partial_z v(Q,z)}{z} \right ]_{z=\epsilon} &=\left.
 \frac{1}{z} \partial_z \left\{ \frac{-1 }{\textit{I}_{0}(Qz_m)} Qz\Bigl( \textit{K}_{0}(Qz_m)\textit{I}_{1}(Q z) - \textit{I}_{0}(Qz_m) \textit{K}_{1}(Q z) \Bigr) \right\} \right |_{z=\epsilon} = \\
&= -\frac{1}{\textit{I}_{0}(Qz_m)}  Q^{2} \Bigl( \textit{K}_{0}(Qz_m) - \textit{I}_{0}(Qz_m) [ln(Q\epsilon/2) + \gamma] \Bigr) = \\
&\xrightarrow[Q \rightarrow \infty]{}  Q^{2} \frac{ln( Q^{2} \epsilon^{2})}{2}\\
\end{align*}
\begin{equation}
\label{17.5}
\Pi_V(Q^{2})= -\frac{R}{g_{5}^{2} Q^{2}} \left [ \frac{\partial_z v(Q,z)}{z} \right ]_{z=\epsilon} = -\frac{R}{2 g_{5}^{2}}ln( Q^{2} \epsilon^{2})
\end{equation}

It was shown in \cite{Erlich,Pomarol1} that this result can be compared with the correlator of vector currents in QCD to obtain the value of $g_5$.
In QCD we have \cite{sumrule}
$$
\Pi_V (Q^{2}) = -\frac{N_c}{24 \pi^{2}} ln Q^{2}
$$
hence $g_5$ is set to
\begin{equation}
\label{18}
\frac{g_{5}^{2}}{R}= \frac{12 \pi^{2}}{N_c}
\end{equation}

\subsection {$\langle J_A,J_A \rangle$ correlator}

Now we shall evaluate the correlator of the axial currents, associated with $A_{\perp}$. It can be evaluated analytically only in the limit of large $Q^{2}$, where we have computed the solution to the equation of motion.

Using the same procedure as for the vector current we obtain the variation of action on the classical trajectories as the boundary term:
\begin{align*}
\delta S_{A_{\perp}} = \left ( - \frac{R}{g_5^{2}} \right ) \int d^{4} x d^{4} q_{1} d^{4} q_{2} e^{ \imath q_{1} x} e^{ \imath q_{2} x} \left [ \delta A_{\perp\mu}(q_{1},z) \frac{ \partial_z A_{\perp\mu}(q_{2},z)}{z} \right ]_{z=\epsilon}.
\end{align*}
It has the same form as for the vector current, so we have
$$
\Pi_A (Q^2)=-\frac{R}{g_{5}^{2} Q^{2}} \left [ \frac{\partial_z a(Q,z)}{z} \right ]_{z=\epsilon}.
$$

\noindent Substituting the solution (\ref{a}), we get
\begin{align*}
\left [ \frac{\partial_z a^{(0)}(Q,z)}{z} \right ]_{z=\epsilon} &= \frac{-1 }{ \textit{I}_{0}(Qz_m)}  Q^{2} \Bigl( \textit{K}_{0}(Qz_m) - \textit{I}_{0}(Qz_m) [ln(Q\epsilon/2) + \gamma] \Bigr )_{z=\epsilon} \xrightarrow[Q \rightarrow \infty]{}  Q^{2} \frac{ln( Q^{2} \epsilon^{2})}{2}\\
\left [ \frac{\partial_z (a^{(1)}+ a^{(2)})}{z} \right ]_{z=\epsilon} &= - Q^{2} \left[ \frac{8}{5} \lambda - 124 \lambda^{2} +\frac{2}{3}\mu \right]
\end{align*}
and finally
\begin{equation}
\label{19}
\Pi_A (Q^2)=-\frac{R}{2 g_{5}^{2}} \left[ ln Q^{2} + \frac{16}{5} \frac{g_{5}^{2} R^2 \Lambda^2 \sigma^{2}}{Q^{6}} - 248 \frac{(g_{5}^{2} R^2 \Lambda^2 \sigma^{2})^2}{Q^{12}} + \frac{8}{3} \frac{g_{5}^{2} R^2 \Lambda^2 \sigma m}{Q^{4}}\right]
\end{equation}
where $\lambda=\frac{g_{5}^{2} R^2 \Lambda^2 \sigma^{2}}{Q^{6}}$ and $\mu=\frac{g_{5}^{2} R^2 \Lambda^2 (2\sigma m)}{Q^{4}}$ are small parameters in the limit $Q^{2}\rightarrow \infty$.

\subsection {$\langle J_{\pi},J_{\pi} \rangle$ correlator}
In this Subsection we will compute the correlator of pseudoscalar currents $J_{\pi}= \bar{q} \gamma_5 q$. To obtain this correlator, we must find out, what field in our theory is dual to $J_{\pi}$. We have pseudoscalar field $\phi =\partial_{\mu} A_{\mu}$, which is connected with axial vector field $A_{\mu}$, dual to the current $J_{A}=\bar{q}\gamma_5 \gamma_{\mu} q$. Now recall, that:
$$
\partial_{\mu}(\bar{q}\gamma_5 \gamma_{\mu} q ) = 2m_{q}(\bar{q}\gamma_5 q )
$$
If we rewrite this expression in terms of our fields, we find after Fourier transform:
\begin{equation}
\label{J_pi}
Q^2 \phi \sim 2m_{q} J_{\pi}
\end{equation}
and we see, that in our theory the pseudoscalar current is dual to the field $\frac{Q^2 \phi}{2 m}$. Hence, to evaluate its correlator, we shall vary the action with respect to $\frac{Q^2 \phi_{0}(q)}{2 m}$

The variation of action of pseudoscalar fields (\ref{S_A}) on classical solutions is:
\begin{align*}
\delta S_{\pi} &=   \int d^{4} x \ \ \frac{R}{g_{5}^{2}} \left[ \delta \partial_{\mu} \phi \frac{\partial_z  \partial_{\mu} \phi }{z} \right]_{z=\epsilon}  - \Lambda^2 R^3 \left[ \delta \pi \frac{v^2}{z^3} \partial_z \pi \right]_{z=\epsilon} \\
&= \int d^{4} x d^{4} q_{1} d^{4} q_{2} e^{ \imath q_{1} x} e^{ \imath q_{2} x} \left( \frac{R Q^2}{g_{5}^{2}} \left[ \delta \phi(q_1,z) \frac{\partial_z  \phi(q_2,z) }{z} \right]_{z=\epsilon}  - \Lambda^2 R^3 m^2 \left[ \delta \pi(q_1,z) \frac{\partial_z \pi(q_2,z)}{z} \right]_{z=\epsilon} \right),
\end{align*}
hence, the pseudoscalar correlator is given by the second variation with respect to $\frac{Q^2 \phi_{0}(q)}{2 m}$ (\ref{phi_pi}):

\begin{align*}
\langle J_{\pi}(q),J_{\pi}(q) \rangle = \frac{4m^2}{Q^4}\left( -\frac{R Q^4}{2 g_{5}^{2}} ln(Q^2 \epsilon^2)+ \frac{R Q^4}{2 g_{5}^{2}} \frac{Q^2}{g_{5}^{2} R^2 \Lambda^2 m^2} ln(Q^2 \epsilon^2) \right).
\end{align*}
In the chiral limit ($m=0$) first term vanishes and we obtain the result:
\begin{equation*}
\langle J_{\pi}(q),J_{\pi}(q) \rangle = 2 \frac{R}{g_{5}^{2}} \frac{1}{g_{5}^{2} R^2 \Lambda^2} Q^2 ln(Q^2 \epsilon^2),
\end{equation*}
which can be compared to QCD \cite{sumrule}:
\begin{equation*}
\langle J_{\pi}(q),J_{\pi}(q) \rangle _{QCD} = \frac{N_c}{16 \pi^2} Q^2 ln(Q^2 \epsilon^2)
\end{equation*}
This comparison and equation (\ref{18}) allow us to fix $\Lambda$:
\begin{equation}
\label{lambda}
\Lambda^2 = \frac{8}{3} \frac{1}{g_{5}^{2}R^2} = \frac{2N_c}{9\pi^2} \frac{1}{R^3}
\end{equation}

\subsection {Evaluation of condensate}
It remained us to fix the relation between $\sigma$ and quark condensate. In QCD one can evaluate condensate as the variation of vacuum energy with respect to quark mass. In dual theory we shall variate the action on the classical solution.
$$
\langle \bar{q}q \rangle = \frac{\delta\varepsilon_{QCD}}{\delta m_q}= \left. \frac{\delta\ S(X_{0})}{\delta m}\right|_{m=0}
$$
The variation of the action (\ref{2}) on the solution (\ref{3}) is
$$
\delta S_{X} = \int d^4 x \left. \frac{R^3}{z^3} \Lambda^2 4 \partial_z X \delta X \right|_{z=\epsilon}= \int d^4 x \left. \frac{R^3}{z^3} \Lambda^2 (m+3\sigma z^2) z \delta m \right|_{z=\epsilon}
$$
therefore quark condensate is related to $\sigma$ as:
\begin{equation}
\label{sigma}
\langle \bar{q}q \rangle = 3 R^3 \Lambda^2 \sigma = \frac{2 N_c}{3 \pi^2} \sigma
\end{equation}
\subsection {"Left-Right" correlator}

Now, from (\ref{17.5}) and (\ref{19}), we can evaluate the quantity $\Pi_{LR} = \Pi_{A} - \Pi_{V}$ and after substituting fixed values of $g_{5}$~(\ref{18}), and $\Lambda$~(\ref{lambda}) we get in the limit of large $Q^2$:
\begin{equation}
\label{LR}
\Pi_{LR}= - \frac{N_c}{12 \pi^2}  \left[ \frac{64}{15} \frac{\sigma^2}{Q^{6}} + 124 \frac{64}{9} \frac{\sigma^{4}}{Q^{12}} - \frac{24}{9} \frac{\sigma m_q}{Q^{4}} \right].
\end{equation}
We see, that in this result there is no factors of $R$, which is by the $AdS/CFT$ \cite{AdS/CFT}\cite{Maldacena} correspondence related to t'Hooft coupling $\lambda'$
\begin{equation}
\label{gym}
\frac{R^{4}}{4 \pi \alpha'^{2} }=\alpha_{s} N_{c}=\lambda'=N_c g_{ym}^2
\end{equation}
that means, that in our result all terms are of order $O(\lambda'^{0})$ in the limit of large $\lambda'$, and we can read out the asymptotic forms of coefficients in
$$
\Pi_{LR}= f(\lambda') \frac{\sigma^{2}}{Q^{6}} + g(\lambda')\frac{\sigma^{4}}{Q^{12}} + \rho(\lambda')\frac{\sigma m_q}{Q^{4}}.
$$
At $\lambda' \rightarrow \infty$ our calculation predicts:
$$
f(\lambda')\sim g(\lambda')\sim \rho(\lambda') \sim \lambda'^{0}
$$
while at weak coupling regime \cite{sumrule}:
$$
\rho(\lambda') \sim \lambda'^{0} \qquad f(\lambda') = -4\pi \alpha_s \sim \lambda'.
$$

As for the dependence on $N_c$, one can see, that provided at large $N_c$ limit $\langle \bar{q}q \rangle \sim N_c$, $\sigma$ (\ref{sigma}) is of order $O(N_c^{0})$, so each term in the result scales as $N_c$ as expected for current correlator.

\section{Conclusion}

From this work we see, that even the simplest AdS/QCD model allows us to compute the current correlators in QCD in the subleading approximation. The "left-right" correlator is proportional to $\langle qq \rangle$ and $m_q$ parameters, which violate the chiral  symmetry spontaneously and explicitly respectively.
In this work we studied the dependance of quantities computed on $N_c$ and AdS radius, connected with $\lambda'$. We reconstruct the power expansion of correlators at large $Q^2$ and found, that to the leading order all terms are of order $O(\lambda'^0)$.

The result of this paper is an argument in favor of the validity of the hard wall AdS/QCD model in description of chiral symmetry breaking in QCD. It provides additional motivation to search for more complicated and suitable models in this branch, moving towards the goal of finding the way to complete description of QCD via its holographic dual.

\section*{Acknowledgements}

I am grateful to Alexander Gorsky for suggesting this problem and general guidance. I also would like to thank Konstantin Zarembo for useful discussions and Thomas Schaefer for important remarks.

This work is partly supported by the Federal Agency for Atomic Energy of Russia, by the Russian President's Grant for Support of Scientific Schools NSh-3036.2008.2, by RFBR grant 07-01-00526 and by the "Dynasty" foundation.
\appendix
\section{Appendix: Computation of the Green function}
Here we will present the derivation of the Green function for modified Bessel equation (\ref{11.5}) appropriate for our problem, which was computed in \cite{Resh, Randall} for Minkovsky momentum.

Green function is defined as:
$$
\left ( -1+\partial_x^{2} - \frac{1}{x} \partial_x \right ) G(x,x') = x \delta(x-x').
$$
The boundary conditions for $G$ should be chosen as follows (\ref{5.5}):
$$
G(x_0, x') = \partial_{x'} G(x, x_m) = 0.
$$
We will find solutions for $x<x'$ and $x>x'$ and then connect them at $x=x'$.

\noindent For $x>x'$ let us define variable $u=max\{x,x'\}$ and at $x \not= x'$
$$
\left ( -1+\partial_u^{2} - \frac{1}{u} \partial_u \right ) G(u) = 0.
$$

\noindent The solution with given boundary conditions is (\ref{11}):
$$
G(u)= const \cdot u(A \textit{I}_{1}(u) + B \textit{K}_{1}(u))
$$
$$
A=\textit{K}_{0}(x_m) \ \ ; \ \ B= -\textit{I}_{0}(x_m)
$$

\noindent For $x<x'$ we set $v=min\{x,x'\}$ and find constants using the boundary condition at $v=x_0 \rightarrow 0$:

$$
G(v)= const \cdot v(C \textit{I}_{1}(v) + D \textit{K}_{1}(v))
$$
$$
\mbox{b.c.: }G(x_0)= const \cdot x_0( C \textit{I}_{1}(x_0) + D \textit{K}_{1}(x_0)) = 0
$$
$$
C=\textit{K}_{1}(x_0) \ \ ; \ \ D=-\textit{I}_{1}(x_0)
$$

\noindent Now we can write the Green function in the whole interval in the form:
$$
G(u,v)=const \cdot u v \cdot [A \textit{I}_{1}(u) + B \textit{K}_{1}(u)][C \textit{I}_{1}(v) + D \textit{K}_{1}(v)]
$$

Integrating the equation for $G(x,x')$ by $x$ at the vicinity of $x'$ and assuming the continuity of Green function, we get the condition for gluing two solutions. Wherefrom we can define $const$:
\begin{gather*}
\partial_u G(u,v)|_{u=x';v=x'} - \partial_v G(u,v)|_{u=x';v=x'} = x'\\
\begin{split}
const \cdot x' x' \Bigl[ \bigl(A \textit{I}_{0}(x') + B \textit{K}_{0}(x')\bigr)&\bigl( C \textit{I}_{1}(x') + D \textit{K}_{1}(x')\bigr)-\\
-&\bigl(A \textit{I}_{1}(x') + B \textit{K}_{1}(x')\bigr)\bigl(C \textit{I}_{0}(x') + D \textit{K}_{0}(x')\bigr) \Bigr] = x'.
\end{split}
\end{gather*}
\noindent Consider this equation at $x' = x_0 \rightarrow 0 $:
$$
const \cdot x_0 \bigl[(AD-BC)  \textit{I}_{0}(x_0) \textit{K}_{1}(x_0) + (BC-AD)\textit{K}_{0}(x_0) \textit{I}_{1}(x_0)\bigr] = 1
$$

$$
const \cdot x_0 \left [ (AD-BC) \frac{1}{x_0} + (BC-AD)\left [ -ln \left ( \frac{x_0}{2} \right ) -\gamma \right ]  \frac{x_0}{2}  \right ] = 1
$$
$$
const = \frac {1}{AD-BC}
$$

\noindent Finally we obtain the Green function:
\begin{equation}
\label{A1}
G(x,x') =  \frac{xx'}{AD-BC} [ A \textit{I}_{1}(u) + B \textit{K}_{1}(u)] [ C \textit{I}_{1}(v) + D \textit{K}_{1}(v)]
\end{equation}

$$
A=\textit{K}_{0}(x_m); \ \  B= - \textit{I}_{0}(x_m); \ \ C=\textit{K}_{1}(x_0)  ; \ \ D= - \textit{I}_{1}(x_0)
$$
\\

\end{document}